\documentclass[journal=nalefd,manuscript=communication]{achemso}

\usepackage{graphicx}

\author{M. D. Schroer}
\author{J. R. Petta}
\email{petta@princeton.edu}
\affiliation{Department of Physics, Princeton University, Princeton, New Jersey 08544}

\title[\texttt{achemso} demonstration]
{Correlating the Nanostructure and Electronic Properties of InAs Nanowires}

\begin{document}
\begin{abstract}
  The electronic properties and nanostructure of InAs nanowires are correlated by creating multiple field effect transistors (FETs) on nanowires grown to have low and high defect density segments. 4.2 K carrier mobilities are $\sim$4$\times$ larger in the nominally defect free segments of the wire. We also find that dark field optical intensity is correlated with the mobility, suggesting a simple route for selecting wires with a low defect density. At low temperatures, FETs fabricated on high defect density segments of InAs nanowires showed transport properties consistent with single electron charging, even on devices with low resistance ohmic contacts. The charging energies obtained suggest quantum dot formation at defects in the wires. These results reinforce the importance of controlling the defect density in order to produce high quality electrical and optical devices using InAs nanowires.
\end{abstract}

Semiconductor nanowires are uniquely suited for studying nanoscale electronics, photonics and basic physics due to the built-in one dimensional confinement of charge carriers. In particular, III-V nanowires are attractive for their electronic properties. Large electron g-factors \cite{PhysRevB.72.201307,Nilsson:2009kq} and small spin-orbit lengths\cite{dhara:121311,fasth:266801} allow spin-dependent effects to be observed at relatively low magnetic fields. Small electron effective masses make it straightforward to grow nanowires with diameters comparable to the Fermi wavelength, resulting in the observation of quantum confinement effects at room temperature\cite{Yu:2003qf}. Mean free paths of over 100 nm have been measured at room temperature \cite{Thelander:2004jk,Chueh:2008pb}, leading to high performance III-V nanowire ballistic transistors. However, as the long history of two-dimensional electron gasses shows, it is only through careful growth optimization and minimization of defects that semiconductor structures may be made clean enough to display new physics and produce the highest performance devices\cite{klitzing:1980,tsui:1982,Pfeiffer:2003qv}.

III-V nanowires are typically grown via the `Vapor-Liquid-Solid' (VLS) method, in which layer by layer growth is directed by a metal nanoparticle.  Originally described in 1964 by Wagner and Ellis for the growth of silicon `nanowhiskers'\cite{wagner:89}, this method has since been applied to grow a large range of materials\cite{Dick2007631,Huang:2001zr,liang:3202,Han:2004mz}. In bulk form, III-V materials occur in either the zincblende or wurtzite crystal structure.\cite{PhysRevB.46.10086} However, for many of these compounds, the energy difference between the two phases is small, on the order of 10 meV/atom for bulk InAs \cite{PhysRevB.52.16936}. In addition, when the dimensions of the system are reduced the surface energy becomes important and can cause the wurtzite phase to be more energetically favorable than zincblende. This transition is predicted to occur at a critical diameter of 10--20 nm in $\left<111\right>$ InAs nanowires.\cite{glas:146101} For slightly larger wires, the structure may alternate between zincblende and wurtzite over nanometer length scales\cite{JJAP.31.2061,Persson:2004kk,Johansson2007635}, resulting in planar defects at the phase boundaries. Recent experiments have connected these defects with reduced carrier lifetime, as determined from photoconductivity measurements, and photoluminescence (PL) intensity.\cite{Joyce:2007gf,Woo:2008qq,Parkinson:2009lq}

The presence of these defects will certainly affect the nanowire's carrier mobility in addition to the optical properties. However, it is not clear that this mobility reduction will be significant when compared with other sources, such as impurity and surface scattering. We address these questions by measuring the mobility in InAs nanowires grown to have segments of both high and low defect density. Nanowires are chosen with a minimal degree of tapering in order to isolate the effects of planar defects on the mobility. We find that the 4.2 K mobility is $\sim$4$\times$ larger in the nominally defect free segments. Coulomb diamonds are observed in finite bias transport measurements taken at low temperatures on high defect density nanowires, even on devices with low resistance contacts, suggesting the formation of quantum dots at defects within the nanowire. We also find a correlation between dark-field optical intensity and mobility, which suggests that simple optical microscopy measurements can be used to sort wires based on their defect density. Our results demonstrate that control of the defect density is of crucial importance if InAs nanowires are to be used as systems for isolation and control of single electron spins.\cite{hanson:1217}

Various methods have been used to control the formation of planar defects, including growing ultrathin nanowires\cite{Shtrikman:2009ud}, using substrates with alternative orientations\cite{krishnamachari:2077,Fortuna:2008fp,Wacaser:2006db}, and modifying the topography of the growth substrate\cite{Shtrikman:2008ly}. However, one of the simplest and most effective methods demonstrated is the two-temperature growth process\cite{Greytak:2004qm,Adhikari:2006xq,Joyce:2007gf}. Here, nanowire growth is initiated during a high temperature nucleation step while subsequent growth of the nanowire is performed at much lower temperatures, producing nanowires with very low levels of defects and tapering.

Nanowires were grown on InAs <111>B substrates using a gold-colloid seeded VLS growth process in a home-built MOVPE growth system.\cite{Schroer:rc} Defect density along the length of the nanowire was controlled using a three step growth process. First, a two min, 500 $^\circ$C growth step nucleates nanowire growth. The temperature was then lowered to 440 $^\circ$C to grow a nominally defect-free segment. Finally, the temperature was raised to 500 $^\circ$C to grow a defect-rich segment. The resulting nanowires ranged from 5 to 12 $\mu$m in length. The growth substrate was sonicated in ethanol to produce a nanowire suspension, which was then dispersed on carbon grids for analysis using a Phillips CM200 transmission electron microscope (TEM). Both bright-field and dark-field TEM show strong banding perpendicular to the growth axis in the section of the nanowire grown during the final high temperature phase (nearest the catalyst particle), as shown in Fig.\ 1. High resolution TEM (HRTEM) shows that the section of nanowire grown during the final high temperature phase contains a high density of defects (>1 nm$^{-1}$), while the section grown at 440 $^\circ$C contains less than one defect per micrometer of length. Additionally, the growth is very uniform with virtually no tapering: the diameter of the nanowires varies less than one nanometer over several micrometers in length. Variations in the nanowire mobility due to a length dependent diameter can be neglected due to the extremely low amount of tapering in these nanowires. To eliminate the possibility of gold contamination from the catalyst particle skewing our results, EDX chemical analysis was performed both in the TEM and a FEI Quanta ESEM with an STEM stage. Within the sensitivity of each tool, the measurements showed no trace of gold in any section of the nanowire other than the catalyst particle.

\begin{figure}
	\centering
	\includegraphics[width=\columnwidth]{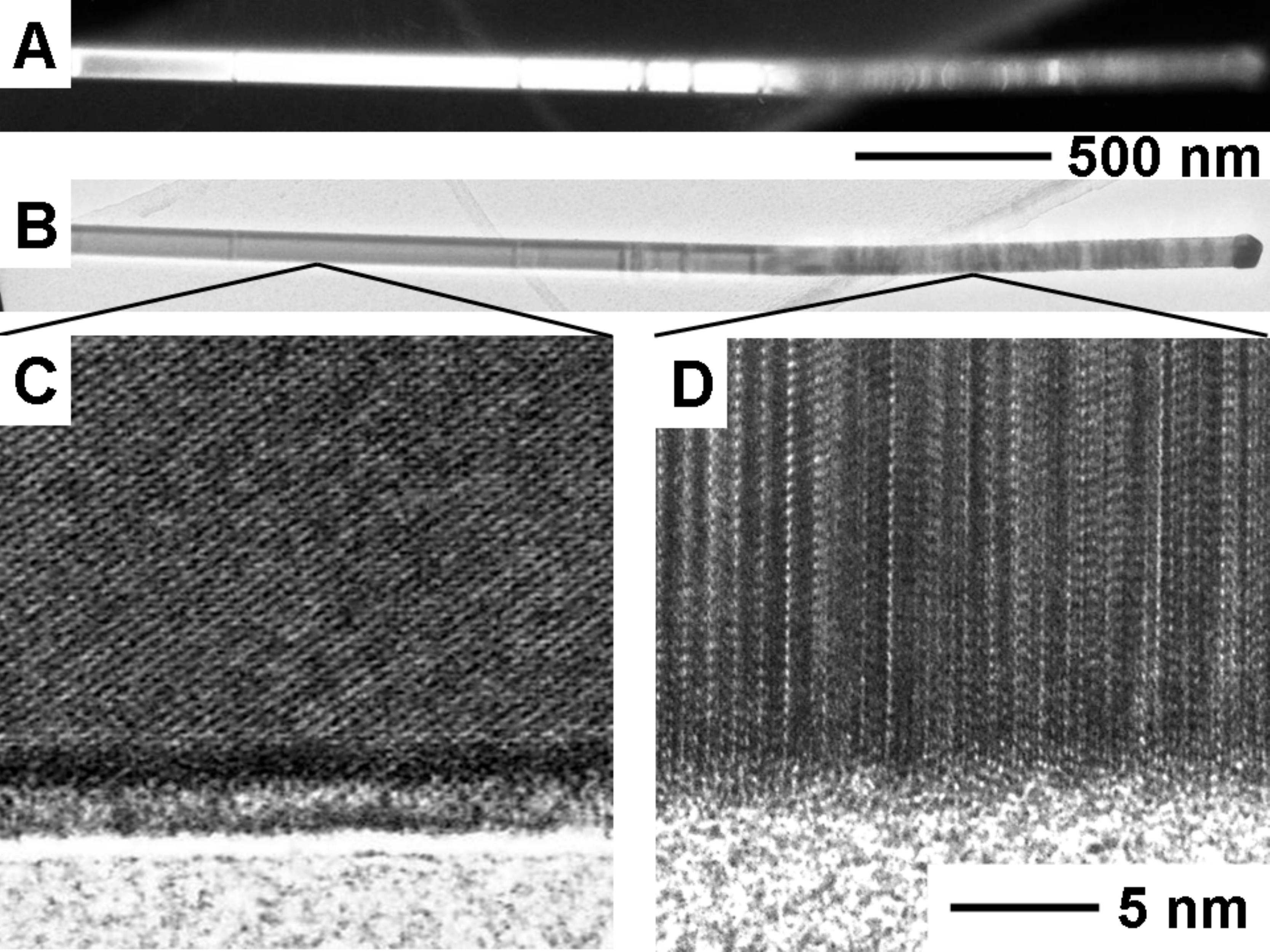}
	\caption{\small TEM images of an InAs nanowire grown using a three step growth process. (a) Dark and (b) bright field TEM images clearly show a largely defect-free region grown at T$_g$ = 440 $^\circ$C followed by a defect rich section of wire nearest the gold catalyst particle (T$_g$ = 500 $^\circ$C). (c) HRTEM of the low defect density region. (d) HRTEM of the high defect density region.}
	\label{fig:TEM}
\end{figure}

Electrical devices were fabricated by dispersing nanowires on silicon substrates covered with 120 nm of thermal oxide. The nanowires were then located relative to pre-patterned alignment marks using dark-field optical microscopy. Twenty four nanowires were selected for device fabrication. Electron beam lithography was used to define a number of contacts along the length of each nanowire, resulting in several field effect transistors (FETs) per nanowire as illustrated in Fig.\ 2a. Each FET was designed to have a contact width of 500 nm and channel length of 1.5 $\mu$m. Ti-Au contacts were thermally evaporated after removing the native oxide on the nanowires using an ammonium polysulfide ((NH$_4$)$_2$S$_x$) etch\cite{0957-4484-18-10-105307}. After liftoff, the samples were annealed at 200 $^\circ$C for one minute. Four probe measurements show that this process produces 4.2 K  contact resistances of 0.5--1 k$\Omega$.

\begin{figure}
	\centering
	\includegraphics[width=\columnwidth]{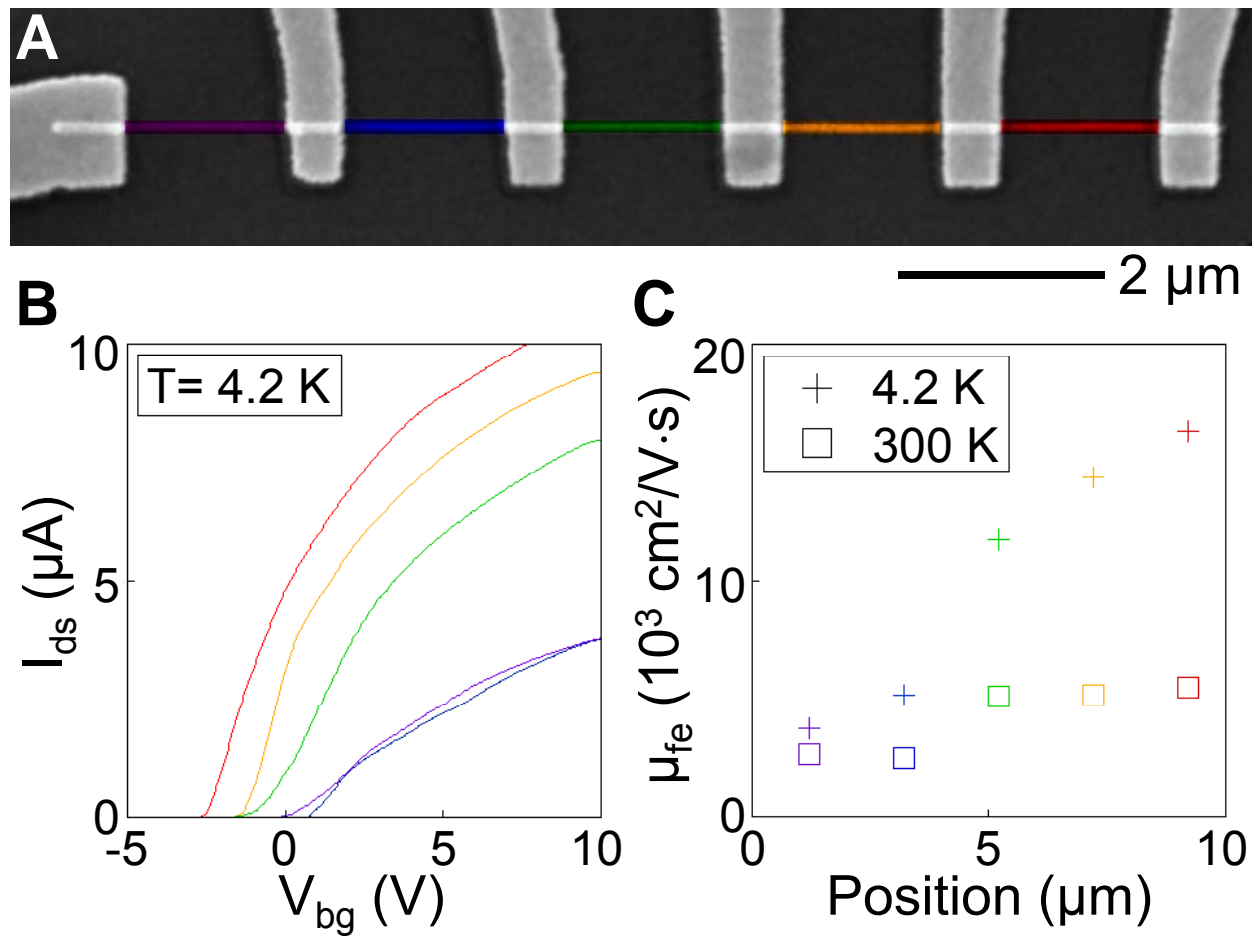}
	\caption{\small Transport properties of a representative nanowire device. (a) False color SEM image of a device with six electrical contacts, resulting in five 1.5 $\mu$m channel length FETs. Each FET is color coded to serve as a key for the data shown in panels b and c. The gold catalyst particle is visible under the contact on the left. The diameter of the nanowire is $\sim$65 nm and is constant along the length of the nanowire to within the measurement uncertainty of the SEM. (b) Current, I$_{ds}$, as a function of backgate voltage, V$_{bg}$, for each of the five FETs shown in panel a. A source-drain bias, $V_{ds}$ = 100 mV, was applied and the backgate voltage was swept at $<$0.1 V/s to reduce hysteresis in the measurements.\cite{Dayeh:2007vf}. (c) Field effect mobilities, $\mu_{fe}$, extracted from the data in panel b.}
\label{fig:transport}
\end{figure}

Two probe transport measurements were performed on each nanowire segment located between a pair of contacts. Figure 2b displays the measured current through each segment as a function of back gate voltage, $I_{ds}$-V$_{bg}$, with a 100 mV source-drain bias, $V_{ds}$. Many of the shorter ($\le$ 6 $\mu$m) devices showed uniform electrical properties over the length of the nanowire, possibly due to cleavage near the middle of the nanowire during sonication. However, five of the longest nanowires showed distinctly different behavior in the two halves of the nanowire. In each case, as the position moves closer to the catalyst particle (and to regions of higher defect density), the on current decreases, the threshold voltage rises and the slope of the traces near depletion decreases. Field effect mobilities were extracted using the I$_{ds}$-V$_{bg}$ transfer curves shown in Fig.\ 2b according to equation 1:

\begin{equation}
	\mu_{fe} = \frac{g_m L_g^2}{ V_{ds} C_g}
	\label{eqn:mobilitycalc}
\end{equation}

where g$_m$ is the peak transconductance, L$_g$ is the gated nanowire length, and C$_g$ is the gate capacitance. The peak transconductance is extracted from transport measurements and the gate length is determined from SEM images of the devices. Directly measuring the gate capacitance for nanowire devices can be challenging, as the extremely low capacitance (on the order of 100 aF) is typically swamped by parasitic capacitances. However, recent work has produced clever means of separating the gate capacitance from the parasitics\cite{Tu:2007xe,Ford:2008qr}. These measurements show that the actual gate capacitance is well approximated by numerical simulations. We therefore used Ansoft's Maxwell finite element solver to calculate the gate capacitances for a range of nanowire diameters.

As may be expected, the regions of the nanowire demonstrating higher conductance correlate with higher field-effect mobilities, as shown in Fig.\ 2c. The room temperature mobilities are about 2$\times$ larger in the nominally defect free segments, in agreement with previous observations\cite{Parkinson:2009lq}. However, the 4.2 K mobility shows a much more pronounced increase of approximately 4.4$\times$. This suggests that while the room temperature mobility is close to being limited by phonon scattering, the planar defects are the dominant factor limiting mobility at low temperatures. The maximum low temperature mobility observed in defect-free segments from various samples ($\sim$16,000--18,000 cm$^2$/V$\cdot$s) compares favorably with InAs/InP core-shell nanowires\cite{Jiang:2007wd}.

\begin{figure}
	\centering
	\includegraphics[width=\columnwidth]{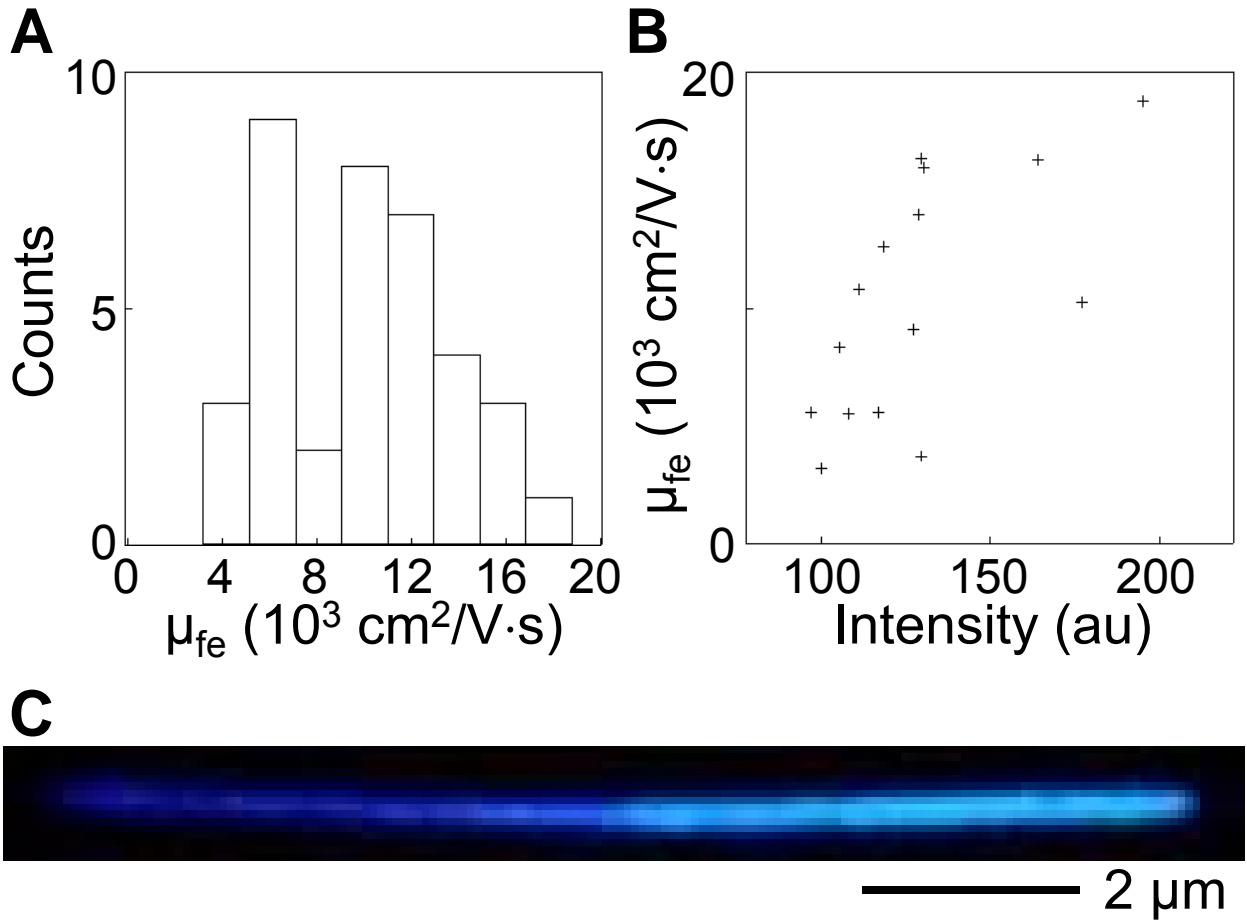}
	\caption{\small (a) Histogram of 4.2 K electron mobilities. (b) 4.2 K electron mobilities versus dark-field optical brightness. The illumination and camera exposure settings were the same for all measurements. (c) Optical image of a nanowire with the catalyst particle on the left, displaying two distinct regions of optical brightness.}
\label{fig:histandbrightness}
\end{figure}

To demonstrate that this is not an isolated effect, a histogram of the 4.2 K mobility from 37 FETs is shown in \ref{fig:histandbrightness}a. Data from all FETs measured are displayed with no attempt to pre-sort individual samples into low or high defect density categories. This histogram appears to show two separate maxima, which are likely attributable to low and high defect density segments. The two maxima are insensitive to how the data are binned. Furthermore, the distribution is consistent with a maximum 4.2 K mobility increase of $\sim$4$\times$ between defect-dense and defect-free regions.

Recent studies have shown that many room temperature optical properties, in particular PL intensity, show degraded values in $\left<111\right>$ oriented III-V nanowires with planar defects.\cite{Parkinson:2009lq,Woo:2008qq,Pemasiri:2009kl}. In the process of locating the nanowires using dark-field optical microscopy, a few nanowires were observed to have two separate regions of high and low optical brightness, as displayed in Fig.\ 3c. Later transport measurements confirmed that these regions are correlated with the high and low mobility sections of the same nanowires. To demonstrate this more fully, Fig.\ 3b shows a plot of measured mobility versus optical brightness for a set of nanowires. These show a reduced set of data versus Fig.\ 3a, due to both the difficulty in assigning regions of the optical image to individual FETs and the relatively large amount of noise present in the optical data.  With this in mind, we produced the data in Fig.\ 3b using one of two methods:  for nanowires which showed homogenous behavior, the highest mobility measured on the wire was plotted versus the optical brightness averaged over a linescan of the entire nanowire. Nanowires which showed a definite change in mobility over the length of the wire were treated as two separate, homogenous nanowires. The resulting graph clearly demonstrates the correlation of mobility with dark-field optical brightness and suggests that the very same optical images used to locate nanowires may be used to easily preselect for the highest mobility nanowires prior to fabrication.

Finally, low bias measurements were performed at 4.2 K. Figure 4a shows two I$_{ds}$-V$_{bg}$ curves, each taken from FETs at opposite ends of a single nanowire. Here, $V_{ds}$ = 1 mV was applied across the FET as $V_{bg}$ was swept from 10 V to -5 V. While the low defect density section of nanowire displays relatively sharp depletion at approximately 0 V, the high defect density segment shows current oscillations well after it appears to be depleted at $\sim$6 V. To further investigate this behavior, we measured the differential conductance (g = dI$_{ds}$/dV$_{ds}$) as a function of back gate voltage, $V_{bg}$, as shown in Figure 4b. These data were taken with an ac excitation of 100 $\mu$V. Peaks in conductance, observed at $V_{ds}=0$, split at finite bias, resulting in `Coulomb diamonds', a signature of single electron charging. However, in contrast with gate defined GaAs quantum dots, the Coulomb oscillations observed here are irregular. The data are suggestive of a multiple quantum dot structure, with each dot having a different characteristic charging energy.

\begin{figure}
	\includegraphics[width=\columnwidth]{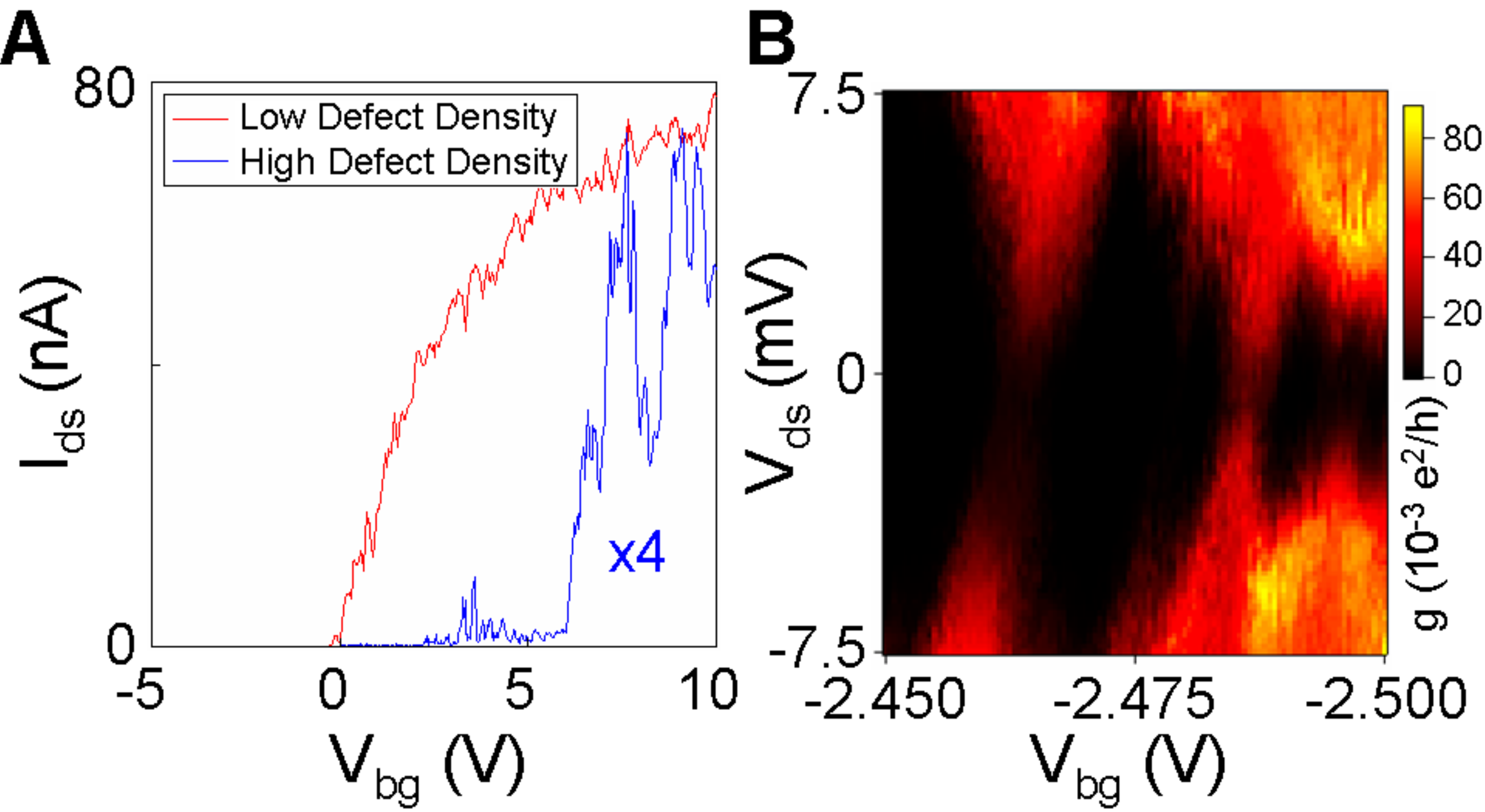}
	\caption{\small Low bias transport at 4.2 K. (a) $I_{ds}$-$V_{bg}$ curves for two FETs located at opposite ends of a single nanowire. (b) Color-scale plot of the nanowire conductance, $g$, as a function of $V_{ds}$ and $V_{bg}$.}
\label{fig:blockade}
\end{figure}

The observation of Coulomb diamonds at 4.2 K is significant, as there is no externally imposed confinement potential. From electrostatics simulations, the total capacitance of a quantum dot consisting of the entire nanowire bounded by the source-drain contacts is 150 aF, which leads to a predicted charging energy of $e^2/(2C)$ = 0.5 meV. The observed charging energies are in the range of 3--7 meV, implying quantum dot sizes of 50--200 nm. Scanned gate microscopy measurements performed by Bleszyncski \textit{et al.}\cite{Bleszynski:2007jw} showed the formation of several series quantum dots in nominally homogenous InAs nanowires. The exact source of the confinement defining the quantum dots was not known, and was attributed to disorder. In the low defect density segments measured here, we find clean depletion with no signatures of Coulomb blockade. These results, combined with previous scanned gate microscopy measurements, indicate that the confinement is being imposed by the defect structure of the nanowire.

Future work may involve performing a more complete measurement of the temperature dependence for both high and low defect density nanowires.  Additionally, measuring the mobility for nanowires with high and low defect density segments of a wide range of diameters could allow the crossover from defect limited mobility to surface scattering mobility to be observed. In practice, this is difficult as the twinning defect density is expected to be dependent on the diameter, but careful calibration of the growth parameters could circumvent this.  Finally, although growing very high quality nanowires has obvious utility, high defect density nanowires may also be useful for investigating various quantum transport phenomenon.  For instance, the influence of a controllable defect density on electron localization could be investigated \cite{Liang:2009qx}.

We have demonstrated a reduction of the twinning defect density in InAs nanowires grown using a two-temperature growth process. Electrical transport measurements on single nanowires containing defect-free and defect-rich segments, illustrate the importance of controlling the twinning defect density in order to fabricate high mobility nanowire devices. At low temperatures, defect rich nanowires showed signatures of Coulomb blockade, most likely due to electronic localization at the defects. These results imply that control over the defect structure could possibly be used to define quantum dots, as an alternative to other methods such as introducing InP tunnel barriers\cite{Fuhrer:2006wj} or through lithography \cite{Fasth:2005wo,Shorubalko:2007le,Shorubalko:2008zt}.   Finally, defect rich nanowires have distinct optical signatures that can be detected using a standard optical microscope, which allows a simple method for pre-selecting the highest mobility wires before device fabrication.

\acknowledgement

We acknowledge support from the Sloan Foundation, the Packard Foundation, the Army Research Office through award W911NF-08-1-0189, and the NSF funded Princeton Center for Complex Materials, DMR-0819860. We also acknowledge the use of the PRISM Imaging and Analysis Center, which is supported in part by the NSF MRSEC program.

\bibliography{nanolett_arxiv_final}

\end{document}